\documentclass[%
reprint,
superscriptaddress,
showpacs,preprintnumbers,
 amsmath,amssymb,	
 aps,
prl,
]{revtex4-1}
\usepackage{graphicx}%
\usepackage{dcolumn}%
\usepackage{bm}%
\pdfoutput=1 
\usepackage{xcolor}
\usepackage{pifont}
\usepackage[]{lineno}
\begin{document}


\title{Study of Two-Photon Exchange via the Beam Transverse Single Spin Asymmetry in Electron-Proton Elastic Scattering at Forward Angles over a Wide Energy Range}

\author{B.~Gou}
\email{boxingou@uni-mainz.de}
\affiliation{%
Institut f\"ur Kernphysik, Johannes Gutenberg-Universit\"at Mainz, J.J. Becherweg 45, D-55099 Mainz, Germany
}%
\affiliation{%
Helmholtz-Institut Mainz, Johannes Gutenberg-Universit\"at Mainz, Staudingerweg 18, D-55099 Mainz, Germany
}%
\affiliation{%
Institute of Modern Physics, Chinese Academy of Sciences, Lanzhou 730000, China
}%
\author{J.~Arvieux} 
\affiliation{%
Institut de Physique Nucl\'{e}aire, CNRS-IN2P3, Universit\'{e} Paris-Sud, F-91406 Orsay Cedex, France
}%
\author{K.~Aulenbacher} 
\affiliation{%
Institut f\"ur Kernphysik, Johannes Gutenberg-Universit\"at Mainz, J.J. Becherweg 45, D-55099 Mainz, Germany
}%
\affiliation{%
Helmholtz-Institut Mainz, Johannes Gutenberg-Universit\"at Mainz, Staudingerweg 18, D-55099 Mainz, Germany
}%
\author{D.~Balaguer~R\'{i}os} 
\affiliation{%
Institut f\"ur Kernphysik, Johannes Gutenberg-Universit\"at Mainz, J.J. Becherweg 45, D-55099 Mainz, Germany
}%
\author{S.~Baunack} 
\affiliation{%
Institut f\"ur Kernphysik, Johannes Gutenberg-Universit\"at Mainz, J.J. Becherweg 45, D-55099 Mainz, Germany
}%
\author{D.~Becker} 
\affiliation{%
Institut f\"ur Kernphysik, Johannes Gutenberg-Universit\"at Mainz, J.J. Becherweg 45, D-55099 Mainz, Germany
}%
\author{L.~Capozza} 
\affiliation{%
Institut f\"ur Kernphysik, Johannes Gutenberg-Universit\"at Mainz, J.J. Becherweg 45, D-55099 Mainz, Germany
}%
\affiliation{%
Helmholtz-Institut Mainz, Johannes Gutenberg-Universit\"at Mainz, Staudingerweg 18, D-55099 Mainz, Germany
}%
\author{W.~Deconinck}
\altaffiliation[Now at ]{Department of Physics and Astronomy, University of Manitoba.}
\affiliation{%
Laboratory for Nuclear Science and Department of Physics, MIT, Cambridge, Massachusetts 02139, USA
}%
\author{J.~Diefenbach} 
\affiliation{%
Institut f\"ur Kernphysik, Johannes Gutenberg-Universit\"at Mainz, J.J. Becherweg 45, D-55099 Mainz, Germany
}%
\author{R.~Frascaria} 
\affiliation{%
Institut de Physique Nucl\'{e}aire, CNRS-IN2P3, Universit\'{e} Paris-Sud, F-91406 Orsay Cedex, France
}%
\author{M.~Gorchtein} 
\affiliation{%
Institut f\"ur Kernphysik, Johannes Gutenberg-Universit\"at Mainz, J.J. Becherweg 45, D-55099 Mainz, Germany
}%
\author{B.~Gl\"{a}ser} 
\affiliation{%
Institut f\"ur Kernphysik, Johannes Gutenberg-Universit\"at Mainz, J.J. Becherweg 45, D-55099 Mainz, Germany
}%
\author{D.~von~Harrach} 
\affiliation{%
Institut f\"ur Kernphysik, Johannes Gutenberg-Universit\"at Mainz, J.J. Becherweg 45, D-55099 Mainz, Germany
}%
\author{Y.~Imai} 
\affiliation{%
Institut f\"ur Kernphysik, Johannes Gutenberg-Universit\"at Mainz, J.J. Becherweg 45, D-55099 Mainz, Germany
}%
\author{E.-M.~Kabu\ss} 
\affiliation{%
Institut f\"ur Kernphysik, Johannes Gutenberg-Universit\"at Mainz, J.J. Becherweg 45, D-55099 Mainz, Germany
}%
\author{R.~Kothe}
\affiliation{%
Institut f\"ur Kernphysik, Johannes Gutenberg-Universit\"at Mainz, J.J. Becherweg 45, D-55099 Mainz, Germany
}%
\author{S.~Kowalski}
\affiliation{%
Laboratory for Nuclear Science and Department of Physics, MIT, Cambridge, Massachusetts 02139, USA
}%
\author{R.~Kunne}
\affiliation{%
Institut de Physique Nucl\'{e}aire, CNRS-IN2P3, Universit\'{e} Paris-Sud, F-91406 Orsay Cedex, France
}%
\author{F.~E.~Maas}
\affiliation{%
Institut f\"ur Kernphysik, Johannes Gutenberg-Universit\"at Mainz, J.J. Becherweg 45, D-55099 Mainz, Germany
}%
\affiliation{%
Helmholtz-Institut Mainz, Johannes Gutenberg-Universit\"at Mainz, Staudingerweg 18, D-55099 Mainz, Germany
}%
\author{H.~Merkel}
\affiliation{%
Institut f\"ur Kernphysik, Johannes Gutenberg-Universit\"at Mainz, J.J. Becherweg 45, D-55099 Mainz, Germany
}%
\author{M.~C.~Mora~Esp\'{i}} 
\affiliation{%
Institut f\"ur Kernphysik, Johannes Gutenberg-Universit\"at Mainz, J.J. Becherweg 45, D-55099 Mainz, Germany
}%
\author{M.~Morlet}
\affiliation{%
Institut de Physique Nucl\'{e}aire, CNRS-IN2P3, Universit\'{e} Paris-Sud, F-91406 Orsay Cedex, France
}%
\author{U.~M\"uller}
\affiliation{%
Institut f\"ur Kernphysik, Johannes Gutenberg-Universit\"at Mainz, J.J. Becherweg 45, D-55099 Mainz, Germany
}%
\author{S.~Ong}
\affiliation{%
Institut de Physique Nucl\'{e}aire, CNRS-IN2P3, Universit\'{e} Paris-Sud, F-91406 Orsay Cedex, France
}%
\author{E.~Schilling} 
\affiliation{%
Institut f\"ur Kernphysik, Johannes Gutenberg-Universit\"at Mainz, J.J. Becherweg 45, D-55099 Mainz, Germany
}%
\author{C.~Weinrich}
\affiliation{%
Institut f\"ur Kernphysik, Johannes Gutenberg-Universit\"at Mainz, J.J. Becherweg 45, D-55099 Mainz, Germany
}%
\author{J.~van~de~Wiele}
\affiliation{%
Institut de Physique Nucl\'{e}aire, CNRS-IN2P3, Universit\'{e} Paris-Sud, F-91406 Orsay Cedex, France
}%
\author{M.~Zambrana}
\affiliation{%
Institut f\"ur Kernphysik, Johannes Gutenberg-Universit\"at Mainz, J.J. Becherweg 45, D-55099 Mainz, Germany
}%
\affiliation{%
Helmholtz-Institut Mainz, Johannes Gutenberg-Universit\"at Mainz, Staudingerweg 18, D-55099 Mainz, Germany
}%
\author{I.~Zimmermann}
\affiliation{%
Institut f\"ur Kernphysik, Johannes Gutenberg-Universit\"at Mainz, J.J. Becherweg 45, D-55099 Mainz, Germany
}%
\affiliation{%
Helmholtz-Institut Mainz, Johannes Gutenberg-Universit\"at Mainz, Staudingerweg 18, D-55099 Mainz, Germany
}%
\date{\today}%

\begin{abstract}
We report on a new measurement of the beam transverse single spin asymmetry in electron-proton elastic scattering, $A^{ep}_{\perp}$, at five beam energies from 315.1~MeV to 1508.4~MeV and at a scattering angle of $30^{\circ} < \theta < 40^{\circ}$. 
The covered $Q^2$ values are 0.032, 0.057, 0.082, 0.218, 0.613~(GeV/c)$^2$.
The measurement clearly indicates significant inelastic contributions to the two-photon-exchange (TPE) amplitude in the low-$Q^2$ kinematic region.
No theoretical calculation is able to reproduce our result. 
Comparison with a calculation based on unitarity, which only takes into account elastic and $\mathrm{\pi N}$ inelastic intermediate states, suggests that there are other inelastic intermediate states such as $\mathrm{\pi \pi N}$, $\mathrm{K \Lambda}$ and $\mathrm{\eta N}$. 
Covering a wide energy range, our new high-precision data provide a benchmark to study those intermediate states.
\end{abstract}

\maketitle

As a probe of hadron structure, electron scattering has two advantages: the structurelessness of the electron and the smallness of the electromagnetic coupling ($\alpha \approx 1/137$).  
The small coupling allows to expand the scattering amplitude in powers of $\alpha$ and to interpret experiments within the one-photon-exchange (Born) approximation. This leading order approximation enables a straightforward extraction of the electromagnetic form factors with the Rosenbluth separation technique~\citep{Rosenbluth1950}. 
For a precise extraction of the form factors it is necessary to include higher order quantum corrections~\citep{Mo:1968cg,Maximon:2000hm}. Importantly, most of those corrections do not alter the Rosenbluth formula in that they contribute an overall factor to the cross section. 

\begin{figure}[b]
\includegraphics[width=0.20\textwidth]{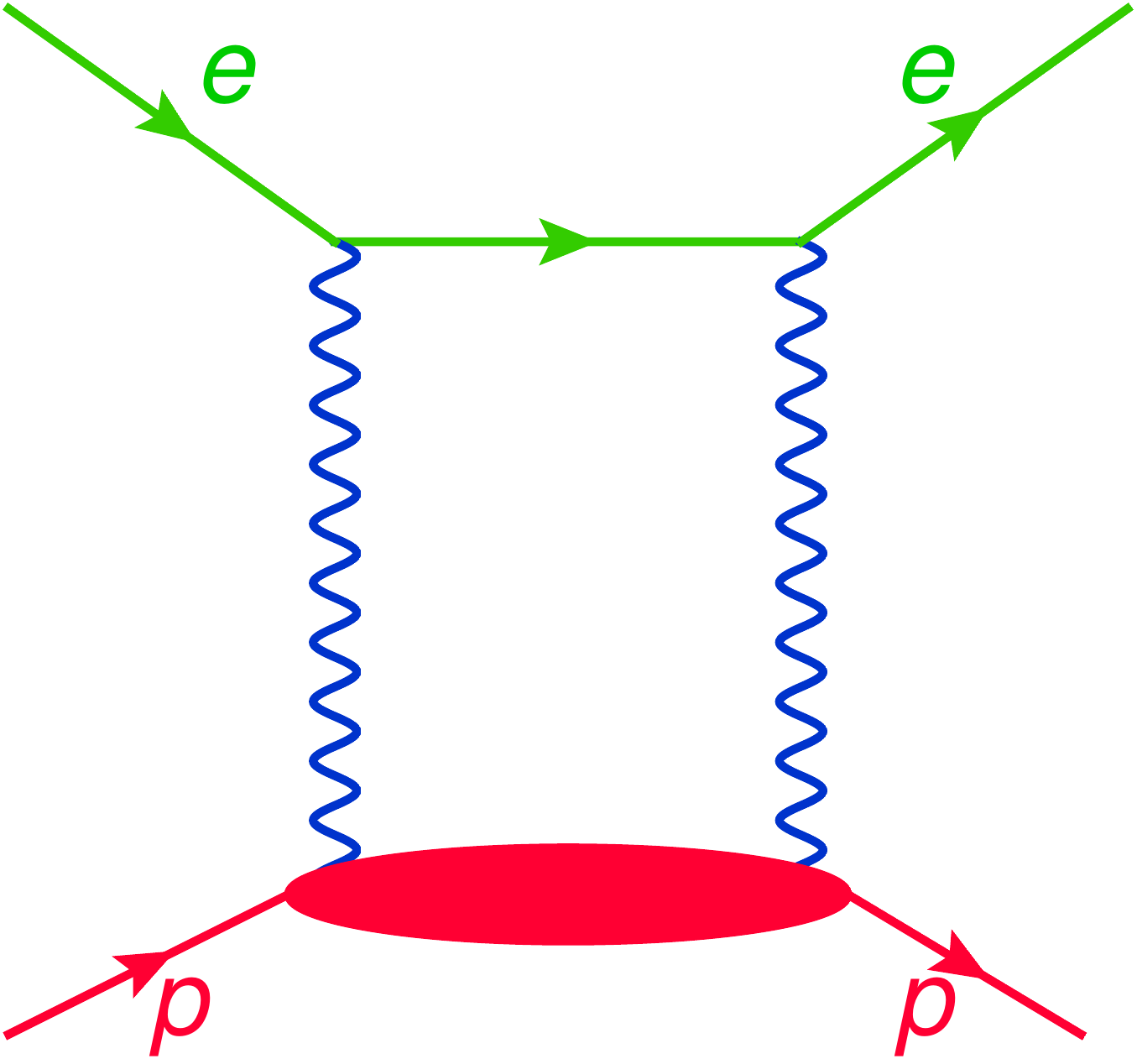}
\caption{\label{fig:TBE} 
The two-photon-exchange contribution to the elastic electron-proton scattering amplitude.
The blob in the lower part represents the doubly-virtual Compton scattering (VVCS) amplitude.}
\end{figure}

The contribution that is expected to break this pattern \citep{Gunion:1972bj,FFPuzzle_TPE_Guichon} is the two-photon-exchange (TPE) diagram depicted in Fig.~\ref{fig:TBE}. For a long time the TPE effects have eluded direct experimental searches~\citep{TPE_Pn_Orsay_1965, TPE_TargetSSA_SLAC_1970, TPE_Rep_DESY_1975}. 
The situation changed when a striking discrepancy between the Rosenbluth separation~\citep{Rexp_Rosenbluth_1994, Rexp_Rosenbluth_2005} and the polarization transfer~\citep{Rexp_Pol_2000, Rexp_Pol_2002,  Rexp_Pol_2010, Rexp_Pol_2011} data on the proton form factor ratio $\mu_pG_E/G_M$ was observed. 
To evaluate the TPE corrections one needs to model the doubly-virtual Compton scattering (VVCS) in the most general kinematics.
This involves calculating the two-current correlator with inclusive hadronic intermediate states. 
The full account of the inclusive intermediate states contribution can be made in the limited near-forward kinematics~\citep{Gorchtein:2014hla}. 
Beyond the forward kinematics, it is only possible to account for the elastic~\citep{FFPuzzle_TPE_Blunden,Borisyuk:2008es,Tomalak:2014dja,Tomalak:2014sva} or the pion-nucleon ($\mathrm{\pi N}$)~\citep{Tomalak:2017shs} intermediate state contributions.
 
The theoretical framework for calculating the TPE contributions plays an important role in evaluating the two-boson-exchange corrections to precision low-energy tests of the Standard Model (SM) in the electroweak sector. 
The proton polarizability contribution to the fine structure of light muonic atoms stems from the TPE diagram and is a substantial ingredient \citep{Carlson:2015jba} in the proton radius puzzle, the $7\sigma$ discrepancy in the value of the proton charge radius extracted from hydrogen spectroscopy \citep{Mohr:2012tt} and electron-proton (ep) scattering \citep{Bernauer:2010wm} on one hand, and muonic hydrogen \citep{Pohl:2010zza,Antognini:1900ns} on the other hand. The situation stays confused with a small proton radius from electron scattering \citep{Xiong:2019umf} and a large proton radius from hydrogen spectroscopy \citep{Fleurbaey:2018fih}.
The hadronic uncertainty of the forward $\gamma Z$-box correction has recently raised a significant interest \citep{Gorchtein:2008px,Sibirtsev:2010zg,Rislow:2010vi,Blunden:2011rd,Gorchtein:2011mz,Blunden:2012ty,Hall:2013hta,Rislow:2013vta,Hall:2015loa,Gorchtein:2015qha,Gorchtein:2015naa,Gorchtein:2016qtl,Erler:2019rmr} in the context of a precision determination of the weak mixing angle with parity-violating electron scattering \citep{Androic:2018kni,Becker:2018ggl}. Similarly, recent works on reducing hadronic and nuclear uncertainties of the $\gamma W$-box correction ~\citep{gW_Marciano_2006,Seng:2018yzq,Seng:2018qru,Gorchtein:2018fxl,Czarnecki:2019mwq} prove central in extracting the $V_{ud}$ element of the Cabibbo-Kobayashi-Maskawa (CKM) matrix and testing the CKM unitarity, a sensitive probe of the SM extensions \citep{Gonzalez-Alonso:2018omy}.
Experimental observables explicitly sensitive to the TPE mechanism are instrumental in developing a dispersion-theoretical framework for TPE.

The lepton-proton scattering amplitude in presence of TPE can be parameterized in terms of six generalized form factors $\tilde{G}_E(Q^2, \varepsilon)$, $\tilde{G}_M(Q^2, \varepsilon)$ and $\tilde{F}_i(Q^2, \varepsilon)$, $i=3, ... , 6$~\citep{SSA_Gorchtein_NPA2004}, where $Q^2$ is the negative four-momentum transfer squared, and $\varepsilon = (1+2(1+\frac{Q^2}{4M^2})\tan^2\frac{\theta}{2})^{-1}$, with $M$ and $\theta$ being the nucleon mass and the lepton scattering angle, respectively. In the Born approximation $\tilde{G}_E(Q^2, \varepsilon)$ and $\tilde{G}_M(Q^2, \varepsilon)$ reduce to the usual electric and magnetic Sachs form factors $G_E(Q^2)$ and $G_M(Q^2)$ which are independent of $\varepsilon$, while the remaining four amplitudes $F_i$ vanish.  
The modified $\varepsilon$ dependence of the transverse and longitudinal transferred polarizations in polarized ep scattering has been investigated~\citep{TPE_Gep2gamma}.
The interference term of the one-photon- and two-photon-exchange amplitudes, which is proportional to the $3^{rd}$ power of the lepton charge, contributes to the cross section with opposite signs in the case of electron-proton and positron-proton scattering. 
One can study the real part of $\tilde{G}_E$, $\tilde{G}_M$ and $\tilde{F}_3$ by measuring the cross section ratio $\frac{\sigma_{e^-p}}{\sigma_{e^+p}}$. 
Such measurements have been performed by CLAS~\citep{TPE_Rep_CLAS_2015}, VEPP-3~\citep{TPE_Rep_VEPP3_2015} and OLYMPUS~\citep{TPE_Rep_OLYMPUS_2017}.

On the other hand, one can study the imaginary part via transverse single spin asymmetries, defined as $A_{\perp}=\frac{\sigma^{\uparrow}-\sigma^{\downarrow}}{\sigma^{\uparrow}+\sigma^{\downarrow}}$, 
where $\sigma^{\uparrow (\downarrow)}$ stands for the cross section with the spin of the polarized particle is parallel (antiparallel) to the normal vector to the scattering plane $\vec{S}_n=\frac{\vec{k}\times \vec{k}^{\prime}}{|\vec{k}\times \vec{k}^{\prime}|}$, with $\vec{k}$ and $\vec{k}^{\prime}$ being the initial and final three-momenta, respectively.
With the spin polarization vector $\vec{P}$ referring to either the polarized target or polarized beam, the asymmetry is expressed as $A_{mea}=A_{\perp}\vec{P} \cdot \vec{S}_n$. This single-spin observable is odd under time reversal~\citep{NSA_NPB_1971}, thus in absence of net $CP$-violation it requires a non-zero imaginary part of scattering amplitudes. The one-photon-exchange amplitude being purely real for spacelike $Q^2$, the transverse spin asymmetry is 
$A_{\perp}=\frac{2Im\mathcal{M}_{2\gamma}\mathcal{M}^{\star}_{1\gamma}}{|\mathcal{M}_{1\gamma}|^{2}}$.
The target asymmetry is sensitive to the imaginary part of $\tilde{G}_E$, $\tilde{G}_M$ and $\tilde{F}_3$ that conserve the lepton helicity, and is of order $O(\alpha)\sim10^{-2}$.
A measurement of the target asymmetry was reported in Ref~\citep{HallA_He3_TPE_2015}.
The beam transverse spin asymmetry is sensitive to the imaginary part of the electron helicity-flip amplitudes $\tilde{F}_{3,4,5}$~\citep{TPE_Pasquini}  
and is of the order of $\alpha \cdot (m_e/E)\sim10^{-6}$ for electron beam energy $E$ in GeV range.
At the time when the interest on the two-photon exchange was revived by the polarization-Rosenbluth discrepancy, the techniques as well as expertise for measuring asymmetries of part per million (ppm) had been developed at several facilities like MIT-Bates, JLab and MAMI, aiming at measuring parity-violating asymmetries in electron scattering~\citep{SFF_Review_Frank}. 
With these facilities, investigations of transverse beam spin asymmetries in various kinematic regions have been performed with transversely polarized electrons scattering off different targets~\citep{SAMPLE_TPE_2001, G0_TPE_2007, G0_TPE_2011, HAPPEX_PREX_TPE_2012, Qweak_TPE_2016, A1_TPE_2018}, including A4 measurements~\citep{A4_TPE_2005, A4_TPE_2017} with both hydrogen and deuterium targets at the A4 experiment.  
In this letter, we report new results of the beam transverse spin asymmetry $A^{ep}_{\perp}$ in ep elastic scattering at forward angles over a wide energy range, measured with a full azimuthal-angle detector at the A4 experiment.

\begin{figure}[]
\includegraphics[width=0.495\textwidth]{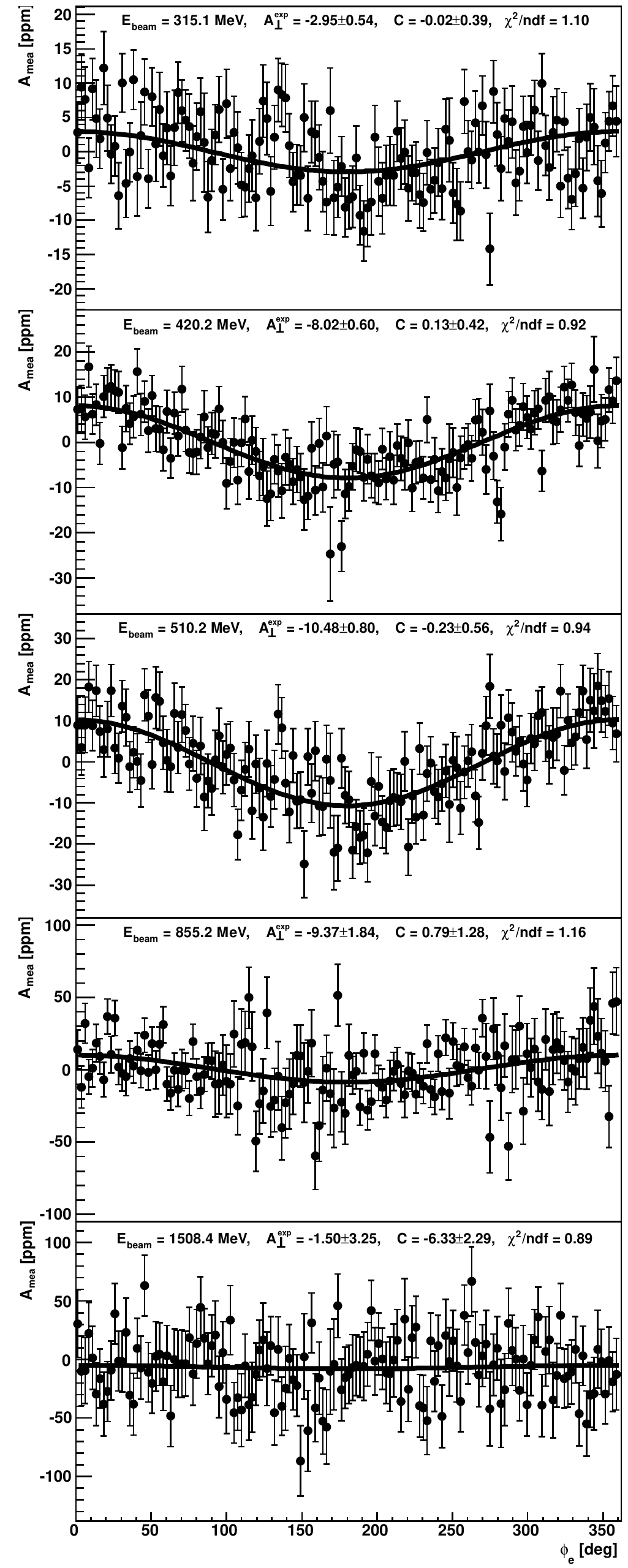}
\caption{\label{fig:AsyFit} 
The asymmetries $\mathrm{A_{mea}}$ measured in 146 frames of the A4 EMC, presented as a function of the electron azimuthal angle $\mathrm{\phi_e}$, fit by $\mathrm{A_{mea}=-A^{exp}_{\perp} cos \phi_e + C}$. 
}
\end{figure}
The experiment was performed at the 1.6~GeV electron accelerator MAMI~\citep{MAMI_EPJA}.
The MAMI electron source provides longitudinally polarized electrons~\citep{MAMI_Pol_Source}, which are produced by illuminating a GaAs super-lattice photocathode with a circularly polarized laser.
A Wien filter is installed in the injection beam line as a spin rotator~\citep{MAMI_Wien}.
The overall effects of the Wien filter and the spin precession in the microtrons lead to a transversely polarized beam at the target position. 
The electron spin is flipped every 20~ms by changing the voltage of the Pockels cell in the laser optics.
To eliminate any possible slow drift effects, the spin flip pattern follows a quadruplet of either $(+--+)$ or $(-++-)$, chosen by a random bit generator.
In order to test and understand any helicity-correlated systematic effects, a half-wave plate, which introduces an extra spin flip, is placed before the GaAs crystal for about $50\%$ of the data-taking time.
MAMI delivers continuous wave polarized electron beams with an intensity of 20~$\mu$A, impinging on a 10~cm long liquid hydrogen target~\citep{A4_Target}.
This gives a luminosity of $\mathrm{L=5.3\times10^{37} cm^{-2} s^{-1}}$, which is monitored with a luminosity monitor (LuMo)~\citep{A4_LuMo}.
The LuMo consists of 8 water Cherenkov detectors and registers scattered electrons emitted at polar angles between $4.4^{\circ}$ and $10^{\circ}$.
To measure the transverse spin asymmetry in ep scattering, 
the electrons scattered between $30^{\circ}$ and $40^{\circ}$ are detected by a fast, totally absorbing, homogeneous electromagnetic calorimeter (EMC) composed of 1022 lead fluoride (PbF$_2$) crystals~\citep{A4_EMC_Cali}. The PbF$_2$ crystals are installed symmetrically about the beam axis in 146 frames, each of which hosts 7 crystals. 
The crystal width is $\frac{4}{3}$ Moli\`{e}re radii ($\mathrm{R_M}$) and the length is larger than 15 radiation lengths ($\mathrm{X_0}$), so more than $95\%$ of an electromagnetic shower are developed in a matrix of $3\times3$ crystals. 
When a valid shower is recognized in a matrix by the self-triggering electronics, the signals from all the 9 crystals are summed and digitized with an 8-bit ADC and stored together with the polarization bit in a histograming unit. 
The histogram is read out and saved on a storage device for each 5-minute run.
The EMC is able to detect electrons with an energy resolution of about $\rm 3.9\%/\sqrt{E/GeV}$, which is sufficient to separate elastically scattered electrons from others.
A typical energy spectrum can be found in Ref.~\citep{A4_PV_2004}.
For every channel there are two such energy spectra, corresponding to the two polarization states ($\uparrow$ and $\downarrow$) respectively.
The number of events for ep elastic scattering is taken as an integral over the elastic peak.
The integral boundaries are carefully determined such that the contamination from inelastic processes is at a negligible level.
Using the numbers of events ($N^{\uparrow}$ and $N^{\downarrow}$) for both polarization states, a raw asymmetry $A_{raw}=\frac{N^{\uparrow} - N^{\downarrow}}{N^{\uparrow} + N^{\downarrow}}$ is obtained for every channel. 
Thanks to the fast spin flip, the systematic effects connected to fluctuations of experimental conditions which are not correlated with helicity, such as the target density, are cancelled out in $A_{raw}$.
However helicity-correlated differences in the beam parameters between the two polarization states could systematically change the measured asymmetry as well. 
For instance, the solid angle covered by a specific crystal is different for two beams with different positions, resulting in a false asymmetry. 
In the same sense, differences in beam angle, beam intensity and beam energy also induce false asymmetries.
In order to correct these false asymmetries, the beam current asymmetry $A_I$, the horizontal and vertical beam position differences $\Delta X$, $\Delta Y$, the horizontal and vertical beam angle differences $\Delta X^{\prime}$, $\Delta Y^{\prime}$, and the beam energy difference $\Delta E$ were measured every 20~ms.
In the offline data analysis, a correction is made for each detector unit, 
i.e. $A_{corr}=A_{raw} - c_1 A_I - c_2 \Delta X - c_3 \Delta Y - c_4 \Delta X^{\prime} - c_5 \Delta Y^{\prime} - c_6 \Delta E$.
The correction coefficients $c_i~(i=1...6)$ are determined through a multiple linear regression analysis. 
After this correction, the asymmetries measured in different data-taking periods, when the half-wave plate was either in or out of the laser optics of the polarized electron source, are consistent with each other, as in our previous investigations~\citep{A4_PV_2004, A4_TPE_2005, A4_PV_2005, A4_PV_2009, A4_PV_2016, A4_TPE_2017}. 
\begin{table*}[t]
\caption{\label{tab:Results}The beam transverse spin asymmetry in electron-proton (ep) elastic scattering ($A^{ep}_{\perp}$) measured at each beam energy. $Q^2$ is determined as cross-section weighted average over the detector acceptance. 
Four systematic uncertainties, which contribute the total systematic error are list in the last rows.}
\begin{ruledtabular}
\begin{tabular}{lccccc}
Beam energy~[MeV] & \phantom{0}315.1 & \phantom{0}420.2 & \phantom{0}510.2 & \phantom{0}855.2 & 1508.4\\
$\mathrm{Q^2}$~[(GeV/c)$^2$] & 0.032 & 0.057 & 0.082 & 0.218 & 0.613\\
$\mathrm{A^{ep}_{\perp}}$~[ppm] & -2.22 & -6.88 & -9.32 & -7.46 & -0.06\\
Statistical error~[ppm] & 0.40 & 0.53 & 0.63 & 1.22 & 2.89\\
Total systematic error~[ppm] & 0.43 & 0.42 & 0.62 & 1.55 & 1.90\\
\colrule
Helicity correlated beam differences & 0.36 & 0.29 & 0.49 & 1.37 & 1.47\\
Polarization measurement & 0.02 & 0.04 & 0.05 & 0.06 & 0.11\\
Spin angle measurement & 0.00 & 0.01 & 0.02 & 0.02 & 0.17\\
Al target window dilution & 0.22 & 0.30 & 0.38 & 0.73 & 1.19\\
\end{tabular}
\end{ruledtabular}
\end{table*}
The corrected asymmetry $\mathrm{A_{corr}}$ is then normalized by the beam polarization $\mathrm{P_e}$, which was measured approximately once per day using a Mott polarimeter located at the beam injection line.
Taking into account the systematic uncertainty of the Mott device and the interpolation of the polarization value between the measurements, we end up with an uncertainty of $\mathrm{\Delta P_e/P_e =4\%}$.
In addition, small corrections due to spin misalignment with respect to the transverse direction at the target position are applied in the data analysis.
Fig.~\ref{fig:AsyFit} shows the normalized asymmetry $\mathrm{A_{mea}=A_{corr}/P_e}$ measured in each frame of the EMC.
By fitting the asymmetry distribution with a function $\mathrm{A_{mea}=-A^{exp}_{\perp} cos \phi_e + C}$ an asymmetry $\mathrm{A^{exp}_{\perp}}$ and an offset C are extracted.
Since the beam current asymmetry has been corrected for, the offset C is a measure for the target density asymmetry. 
Continued improvements of the liquid hydrogen target operation and the beam stabilization systems over 10 years has led to an operation point with reduced target density fluctuation by a factor of 20 as compared to the early measurement~\citep{A4_PV_2004}. Our analysis has shown that we do not need any correction for target fluctuation for the data presented here.
The vanishing offsets measured at 315.1~MeV, 420.2~MeV, 510.2~MeV and 855.2~MeV demonstrate that the target density fluctuation in our experiment was very well controlled. 
The offset measured at 1508.4~MeV deviates from zero by 6.33~ppm, but is compatible with zero within 3~$\sigma$.
The asymmetry $\mathrm{A^{exp}_{\perp}}$ is given by the ep scattering asymmetry $\mathrm{A^{ep}_{\perp}}$, diluted by the background asymmetry $\mathrm{A^{eAl}_{\perp}}$ from the electron-aluminium (eAl) scattering at the target window.
The aluminium dilution factor $\mathrm{f}$, defined as $\mathrm{f=\frac{Y_{eAl}}{Y_{ep}+Y_{eAl}}}$, with $\mathrm{Y_{ep}}$ and $\mathrm{Y_{eAl}}$ being the yield of ep and eAl scattering respectively, was measured to be 0.060 with a relative error of $\mathrm{10\%}$.
For the eAl asymmetry we adopt the theoretical calculation in Ref.~\citep{TPE_Gorchtein_2008}.
The ep asymmetries determined as $\mathrm{A^{ep}_{\perp} = (A^{exp}_{\perp}-f  A^{eAl}_{\perp})/(1-f)}$ are given in Table~\ref{tab:Results}.
The statistical and total systematic uncertainties, as well as uncertainties due to helicity-correlated false asymmetries, beam polarization, spin angle and target window dilution are also listed.

\begin{figure}[t]
\includegraphics[width=0.48\textwidth]{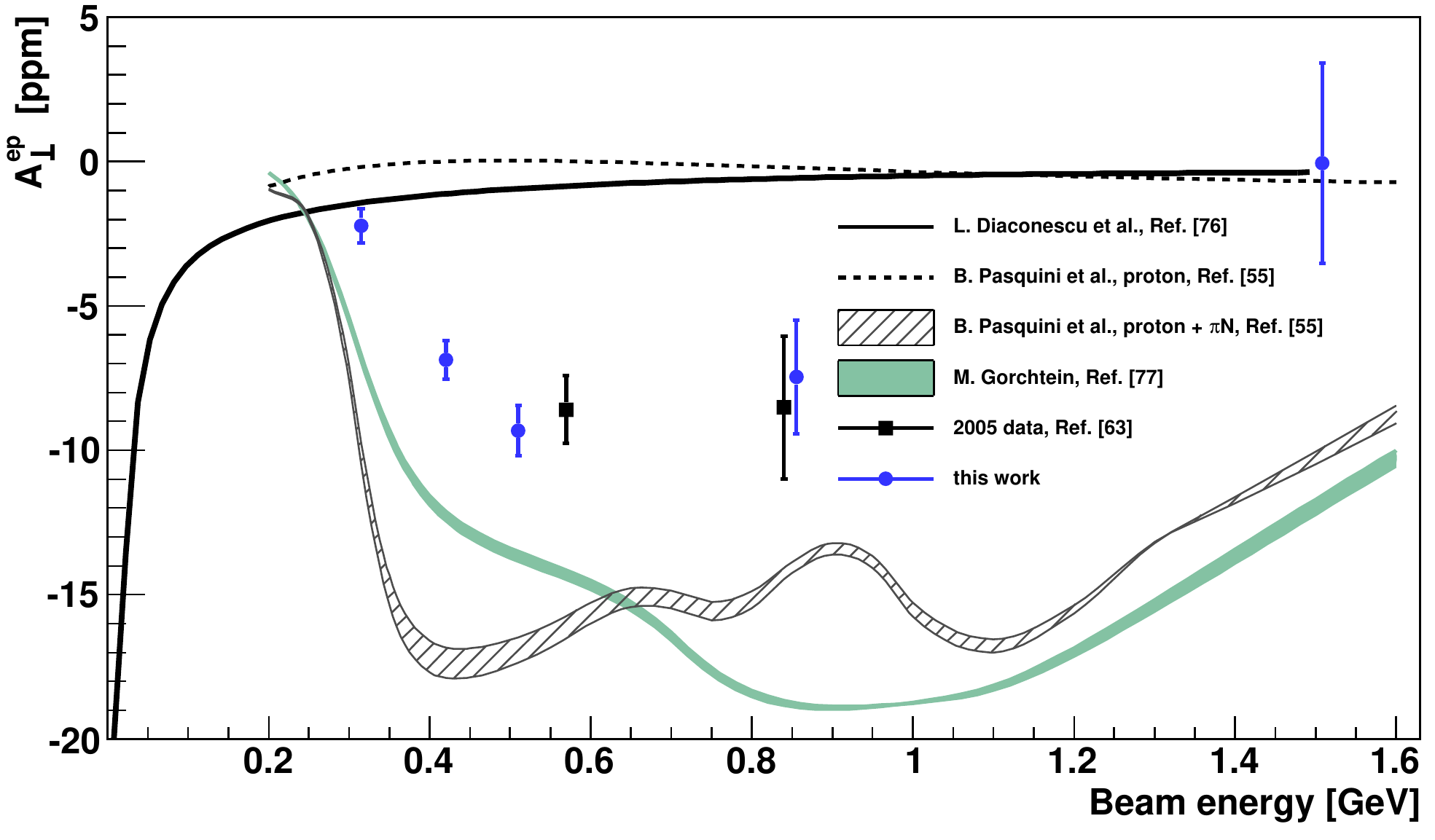}
\caption{\label{fig:A_Ebeam} $A^{ep}_{\perp}$ as a function of beam energy. 
Note the data point at 855.2~MeV from Ref.~\citep{A4_TPE_2005} is shifted horizontally for a better view.
The bands reflect the dependence of the mean scattering angle on beam energy.
}
\end{figure}

As shown in Fig.~\ref{fig:A_Ebeam}, the ep beam transverse spin asymmetries ($A^{ep}_{\perp}$) measured in this work are consistent with our previous measurements~\citep{A4_TPE_2005}. 
More importantly, the new measurements substantially expand the energy range, thus enable TPE studies in a vastly extended kinematic region. 
Our experimental data show that $A^{ep}_{\perp}$ increases with beam energy from 315.1~MeV to 510.2~MeV, and reaches a plateau between 510.2~MeV and 855.2~MeV. 
At the first three energies our experimental errors are smaller than 1~ppm.
With the decrease of the ep scattering cross section, the measurements at higher energies have increased statistical errors.
Despite of the large uncertainty, the asymmetry measured at 1508.4~MeV is consistent with zero.
To understand the data, several theoretical calculations are shown in Fig.~\ref{fig:A_Ebeam} as well.
The heavy baryon chiral perturbation theory adopted in Ref.~\citep{VAP_Diaconescu} (solid black curve) is seen to reproduce our data point at 315.1~MeV but is only valid at much smaller incident electron energy. 
In Ref.~\citep{TPE_Gorchtein} the imaginary part of the VVCS amplitude is related to the total photoabsorption cross section $\sigma_{\gamma p}$ by the optical theorem. While optical theorem is only applicable in the exact forward limit, Ref.~\citep{TPE_Gorchtein} proposed a phenomenological approach to extend it to small finite values of $Q^2$. The updated calculation in the relevant kinematics in that approach is represented by the green curve. 
Ref.~\citep{TPE_Pasquini} accounts for the elastic and $\mathrm{\pi N}$ intermediate states. 
The dashed line represents the elastic contribution, which is expressed in terms of the proton from factors $G_E$ and $G_M$.
The $\mathrm{\gamma^{\star}p \rightarrow \pi N}$ amplitudes are taken from the latest MAID analyses of single $\pi$ electroproduction observables~\citep{MAID_2007}.
This calculation has given results which agree well with our backward-angle data~\citep{A4_TPE_2017}. 
As shown in Fig.~\ref{fig:A_Ebeam}, our forward-angle asymmetries are significantly smaller than both calculations. 
The substantial discrepancy might be resolved by including higher-mass intermediate states such as $\mathrm{\pi \pi N}$, $\mathrm{K \Lambda}$ and $\mathrm{\eta N}$ in the calculation of Ref.~\citep{TPE_Pasquini}. For the calculation in Ref.~\citep{TPE_Gorchtein}, off-forward contributions need to be added.  
Covering a broad range of exchanged photon energies and virtualities, our measurements offer possibilities to benchmark future extensions of theoretical calculations. 
We acknowledge the crew of the MAMI accelerator
for the high beam quality. 
We thank B.~Pasquini for the theoretical calculation.
This work has been supported by the Deutsche
Forschungsgemeinschaft (DFG) within the projects
SFB 443, SFB 1044, and the PRISMA excellence cluster.
The author B.~Gou would like to thank the Office of China 
Postdoctoral Council (OCPC) for their partial financial support.

\bibliography{ms}
\end{document}